\documentclass{PoS}

\title{Radiative corrections to neutralino annihilation: Recent developments}
\ShortTitle{Radiative corrections to dark matter annihilation: Recent developments}

\author{\speaker{Bj\"orn Herrmann}\\
        Deutsches Elektronen-Synchrotron (DESY), Notkestra{\ss}e 85, D-22603 Hamburg, Germany\\
        E-mail: \email{bjoern.herrmann@desy.de}}

\abstract{Evaluating the relic density of dark matter is an interesting possibility to constrain the parameter space of new physics models. However, this calculation is affected by several sources of uncertainty. On the particle physics side, considerable progress has been made in the recent years concerning the calculation of the annihilation cross-section of dark matter, which is needed in this context. In particular, within the Minimal Supersymmetric Standard Model, the theoretical uncertainty has been reduced through the calculation of loop corrections. The present contribution gives an overview over the achievements that have been made in QCD corrections to neutralino pair annihilation. The numerical impact is illustrated for a few examples.}

\FullConference{Identification of Dark Matter 2010-IDM2010\\
		July 26-30, 2010\\
		Montpellier, France}

\begin{document}

% =======================================================================
\section{Motivation}

Among the most compelling evidence for physics beyond the Standard Model of particle physics is the presence of cold dark matter (CDM) in our universe. Assuming the standard cosmological $\Lambda$CDM scenario, recent cosmological measurements constrain its relic density to
\begin{equation}
	0.1053 ~<~ \Omega_{\rm CDM}h^2 ~<~ 0.1193
\label{eq:omh2}
\end{equation}
at 95\% confidence level. The given limit is based on the 7-year data from the WMAP mission in combination with other observations \cite{WMAP}. %Soon, there will be even more precise cosmological measurements from the PLANCK satellite.
%Let us already note that there will be even more precise limits from the PLANCK satellite in a near future.

For a given new physics scenario, the relic abundance of a dark matter candidate in our universe can be evaluated by solving the Boltzmann equation
\begin{equation}
	\frac{{\rm d}n}{{\rm d}t} ~=~ -3 H n - \langle\sigma_{\rm ann}v\rangle \left( n^2 - n_{\rm eq}^2 \right),
\label{eq:boltzmann}
\end{equation}
which describes the evolution of the number density $n$ of the dark matter candidate. The first term on the right hand side conveys the dilution due to the expansion of the universe, $H$ denoting the (time-dependent) Hubble parameter. The second term is related to annihilation and co-annihilations of dark matter. Here, $n_{\rm eq}$ is the number density in thermal equilibrium. The total (co-)annihilation cross-section $\sigma_{\rm ann}$, multiplied by the relative velocity $v$, has to be convolved with the velocity distribution to obtain the thermal averaged cross-section $\langle \sigma_{\rm ann} v \rangle$. After integrating the Boltzmann equation, Eq.\ (\ref{eq:boltzmann}), the relic density of dark matter is given by $\Omega_{\rm CDM}h^2 = m_{\chi}n_0/\rho_{c}$, $m_{\chi}$ and $n_0$ being the mass and present number density of the dark matter particle, respectively, and $\rho_c$ the critical density of the universe. This cosmological constraint gives additional information with respect to mass limits from collider searches and precision measurements such as the rare decay $b\to s\gamma$ or the anomalous magnetic moment of the muon $(g-2)_{\mu}$.

In the Minimal Supersymmetric Standard Model (MSSM), one of the most convincing dark matter candidates is the lightest neutralino, which is stable if it is the lightest superpartner and if $R$-parity is conserved. Several public tools exist, that perform a numerical calculation of the neutralino relic density for a given set of parameters. In these packages, such as {\tt DarkSUSY} \cite{DarkSUSY} or {\tt micrOMEGAs} \cite{micrOMEGAs}, all processes that contribute to the annihilation cross-section are implemented, but mostly only as a pure leading order calculation. Radiative corrections are included only for very sensitive quantities, e.g., the bottom Yukawa coupling. However, it is well known that next-to-leading order corrections can have sizeable impacts on cross-sections, and may in consequence alter the prediction of the dark matter relic density.

Recently, several results for corrections to the dark matter annihilation cross-section have been published. Here, we focus on corrections of ${\cal O}(\alpha_s)$ \cite{AFunnel, DM_mSUGRA, DM_NUHM}, which are most important in case of neutralino pair annihilation into quarks. For other processes, such as annihilation into gauge bosons or leptons, electroweak corrections can be important in the same manner. Details and a numerical analysis for this case can be found in Refs.\ \cite{BaroNLO}. Moreover, one-loop corrections due to the exchange of a gauge or Higgs boson in the initial state have been investigated in Ref.\ \cite{Drees}. Finally, let us note that the prediction of the relic density is affected by further uncertainties. 

In the next Section, a brief overview over the annihilation of neutralinos into quarks and the associated corrections of order $\alpha_s$ is presented. Numerical examples follow in Sec.\ \ref{sec3}. Finally, conclusions are given in Sec.\ \ref{sec4}.

% =======================================================================
\section{Neutralino annihilation into quark-antiquark pairs \label{sec2}}

\begin{figure}
	\begin{center}
	\includegraphics[scale=1.0]{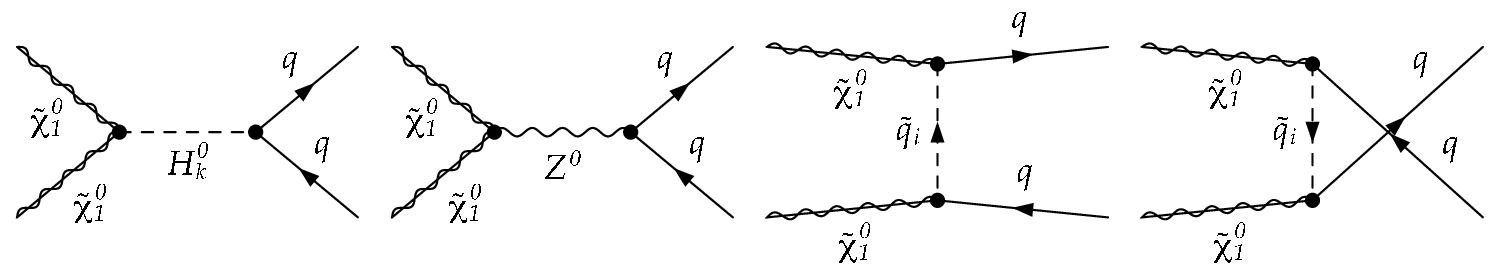}
	\end{center}
	\vspace{-5mm}
	\caption{Tree-level processes contributing to neutralino pair annihilation into quark-antiquark pairs. Annihilation is possible through the exchange of a Higgs-boson ($H_k^0=h^0,H^0,A^0$), a Z-boson, or a squark ($\tilde{q}_i = \tilde{q}_1, \tilde{q}_2$).}
	\label{fig:treelevel}
\end{figure}

In the following, we focus on neutralino pair annihilation into quark-antiquark pairs. The relevant Feynman diagrams are depicted in Fig.\ \ref{fig:treelevel}. In the constrained MSSM (cMSSM) with unification conditions at the GUT scale, the annihilation into third generation quarks dominates in wide ranges of the WMAP-favoured parameter space \cite{DM_mSUGRA}. At low values of the gaugino mass parameter $m_{1/2}$, the annihilation proceeds efficiently through a resonance of the light scalar Higgs-boson $h^0$ into $b\bar{b}$ final states. In the so-called focus point region at high scalar masses $m_0$, important final states are $t\bar{t}$ that occur through the resonance of the heavy scalar Higgs $H^0$. Finally, for large $\tan\beta$, the so-called ``A-Funnel'' region with resonant annihilation into $b\bar{b}$ through a pseudoscalar Higgs $A^0$ opens.

While in the cMSSM, mostly Higgs-resonances are relevant, the exchange of a Z-boson or a squark can be enhanced if (at least some of) the unification conditions are relaxed. In consequence, the higgsino fraction of the neutralino can be altered, leading to enhanced couplings to the Z-boson. This is the case, e.g., for scenarios with non-universal Higgs masses (NUHM) or non-universal gaugino masses. 
%In the following we will mainly consider only the former, bearing in mind that the results are qualitatively similar in the latter case \cite{DM_NUHM}. 
The exchange of squarks is preferred for light squark masses, which occurs in particular for a large trilinear coupling $A_0$.

The annihilation cross-section related to the diagrams in Fig.\ \ref{fig:treelevel} receives corrections of order $\alpha_s$ due to one-loop diagrams with gluons and gluinos. These include self-energies for the quarks and squarks, corrections to the vertices involving quarks or squarks, and several box diagrams. At the same order in $\alpha_s$, the emission of real gluons from one of the (s)quarks that are present in the tree-level diagrams has to be taken into account. For the full set of virtual and real contributions the reader is referred to Refs.\ \cite{AFunnel, DM_mSUGRA, DM_NUHM}.

All relevant amplitudes have been calculated analytically in Refs.\ \cite{AFunnel, DM_mSUGRA, DM_NUHM}. After renormalization, the virtual part is ultraviolet-finite, the remaining infrared singularities cancel when taking into account the real gluon emission. The obtained results have been implemented in a numerical package, which can serve as extension for the public relic density tools {\tt DarkSUSY} \cite{DarkSUSY} and {\tt micrOMEGAs} \cite{micrOMEGAs}. Note that we also take into account important effects due to the resummation of the Yukawa coupling, in particular running quark masses and the resummation of terms proportional to $\tan\beta$ or the trilinear couplings $A_b$. Details on the calculation as well as concerning the renormalization procedure and the numerical implementation using the dipole subtraction formalism, can be found in Refs.\ \cite{AFunnel, DM_mSUGRA, DM_NUHM} and the references therein.

% =======================================================================
\section{Numerical examples \label{sec3}}

\begin{figure}
	\begin{center}
	\includegraphics[scale=0.36]{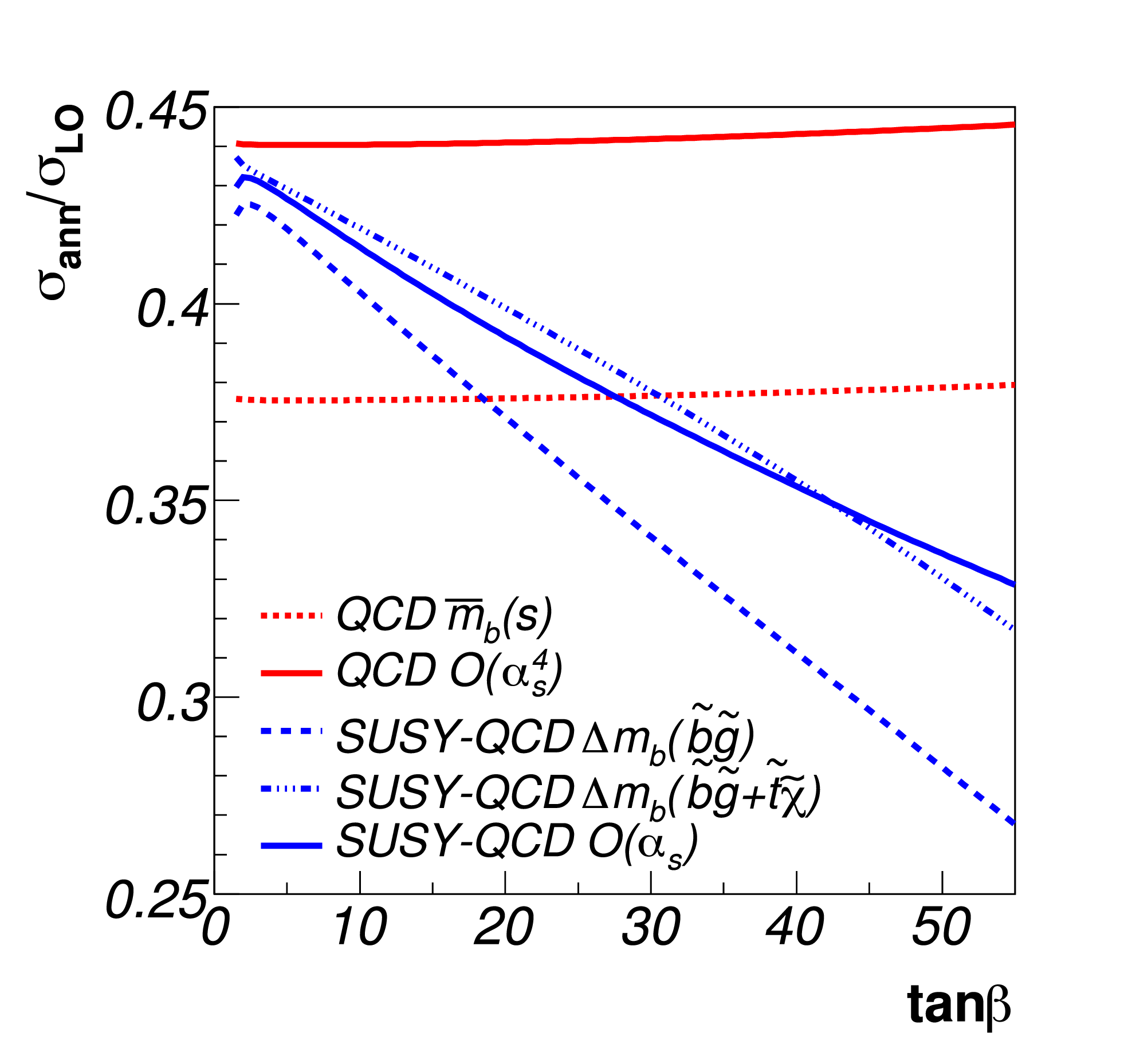}\qquad
	\includegraphics[scale=0.35]{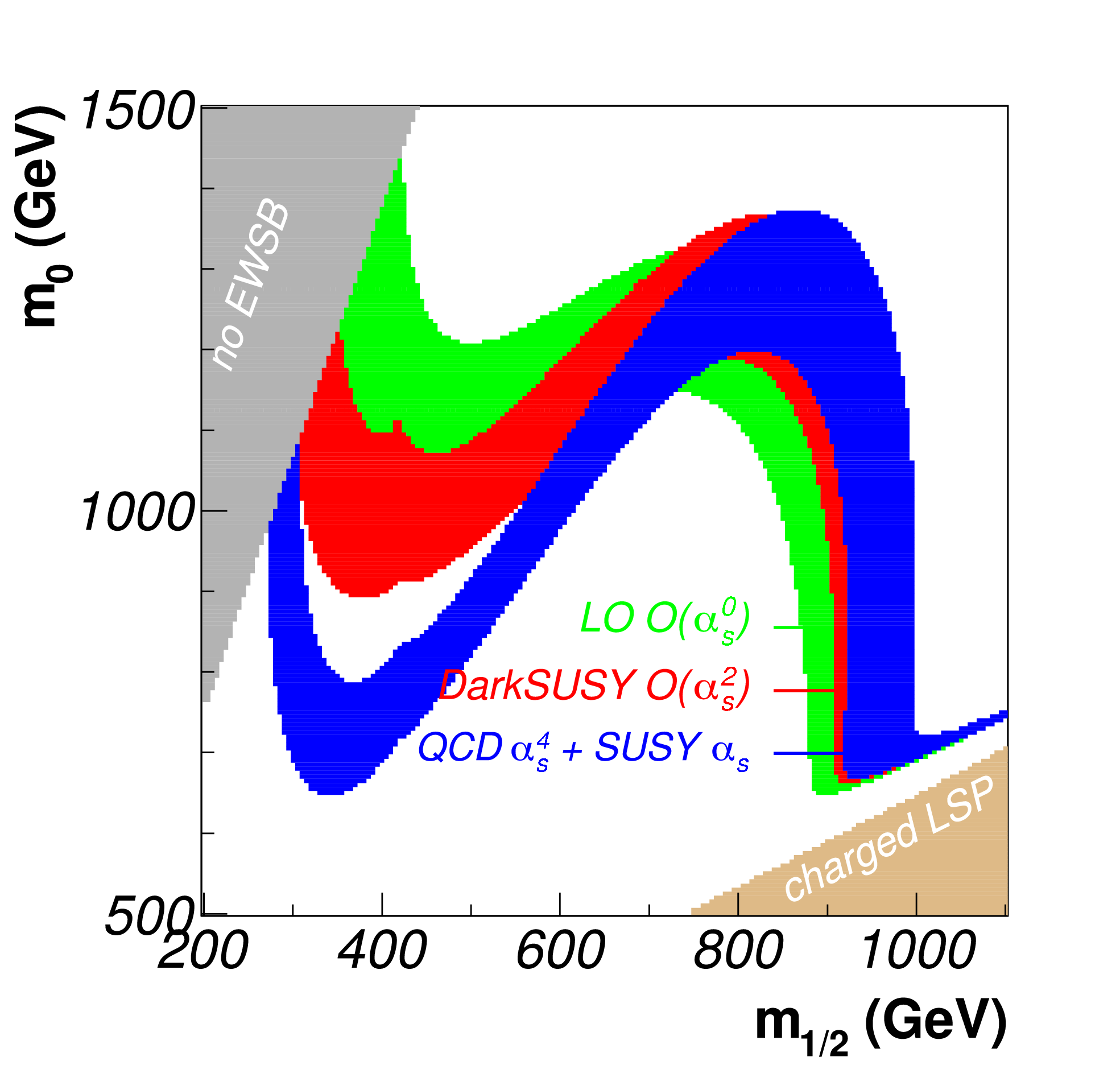}
	\end{center}
	\vspace{-5mm}
	\caption{Impact of the next-to-leading order corrections on the annihilation cross-section for a typical scenario in the ``A-Funnel'' region (left) and the prediction of the neutralino relic density in the $m_0$--$m_{1/2}$ plane for $\tan\beta=54$, $A_0=0$, and $\mu>0$ (right). The shown contours correspond to $0.094 \le \Omega_{\rm CDM}h^2 \le 0.136$.}
	\label{fig:afunnel}
\end{figure}

In the following, the numerical effect of the corrections discussed above is illustrated for a few examples. First, we consider the case of large $\tan\beta$ within the constrained MSSM. In the left panel of Fig.\ \ref{fig:afunnel} we show the annihilation cross-section given by the diagram $\tilde{\chi}^0_1 \tilde{\chi}^0_1 \to A^0 \to b\bar{b}$ including different corrections for a typical reference point in the WMAP-preferred region of parameter space. The numerical values are normalized to the cross-section at leading order. The QCD corrections decrease the cross-section by more than 60\%, which is mainly an effect of the running bottom-quark mass. Finite terms up to ${\cal O}(\alpha_s^4)$ contribute by about 5\%. 

The numerical impact of the SUSY-QCD corrections strongly increases with $\tan\beta$. While the effect is almost negligible for small $\tan\beta \sim 2$, the cross-section is decreased by another 10\% for $\tan\beta \gtrsim 50$, which is the case for the A-Funnel region. Again, the main impact comes from the resummation of the bottom Yukawa coupling, while finite terms are numerically less important \cite{AFunnel}. 

On the right panel of Fig.\ \ref{fig:afunnel}, we show the WMAP-favoured region of the $m_0$--$m_{1/2}$ plane for fixed $\tan\beta = 54$, $A_0 = 0$, and $\mu>0$ and for different levels of correction. The relic density has been evaluated using {\tt DarkSUSY 4.1} \cite{DarkSUSY}, which includes by default the QCD running quark mass and finite terms up to ${\cal O}(\alpha_s^2)$ in the bottom Yukawa coupling, but no SUSY-QCD corrections. The latter have been implemented into the package in order to evaluate their numerical impact. For reference, we also show the prediction of the relic density including a pure tree-level calculation of the annihilation cross-section. As can be seen, the impact of the additional corrections is larger than the uncertainty of the WMAP measurement, especially for low mass parameters. Interestingly, the effect is reversed around the Higgs-pole at $m_{1/2} \sim 800$~GeV. This is explained by the fact, that the decay width of the Higgs-boson receives the same corrections as the annihilation cross-section \cite{AFunnel}.

\begin{figure}
	\begin{center}
	\includegraphics[scale=0.35]{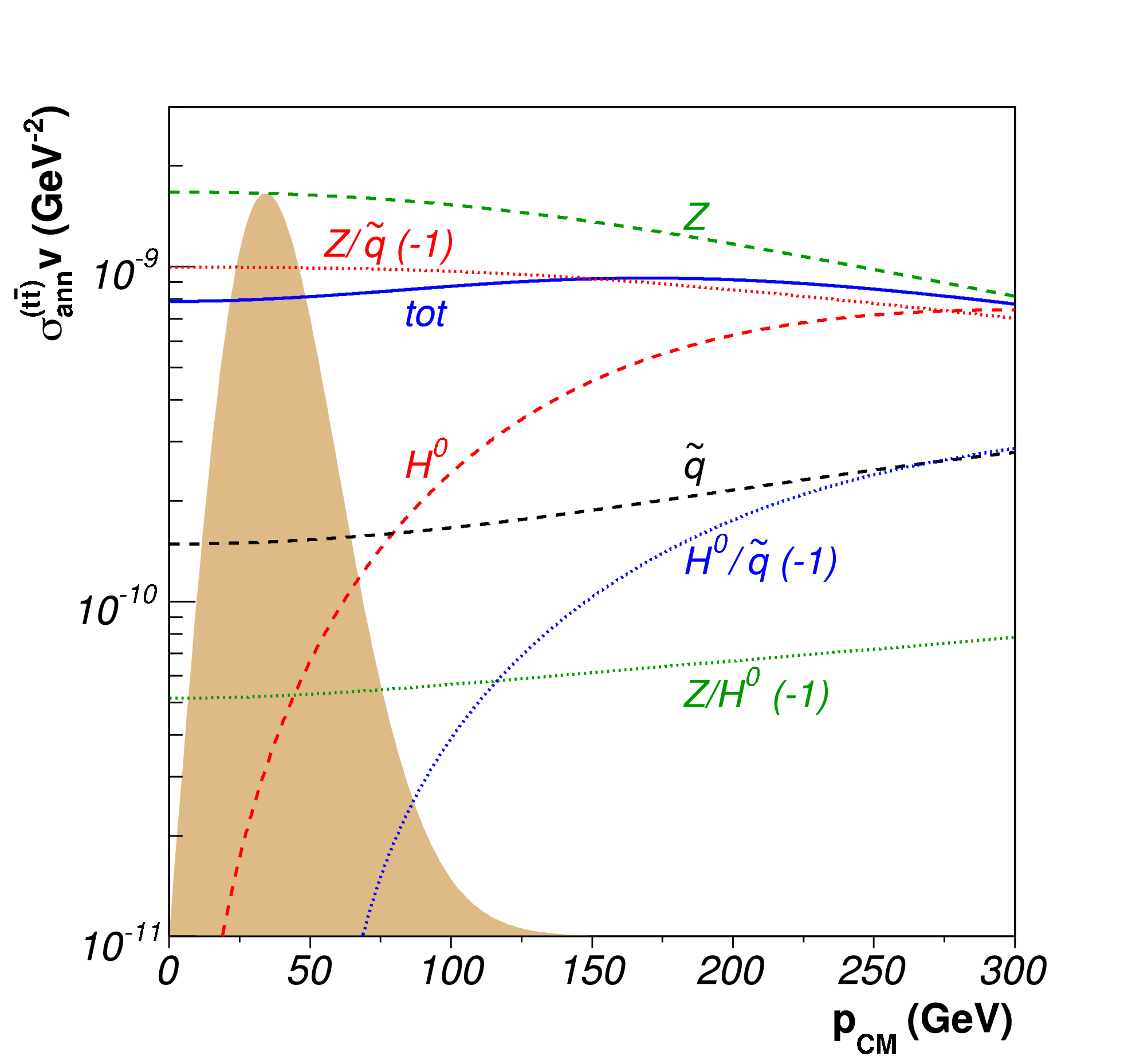}\qquad
	\includegraphics[scale=0.35]{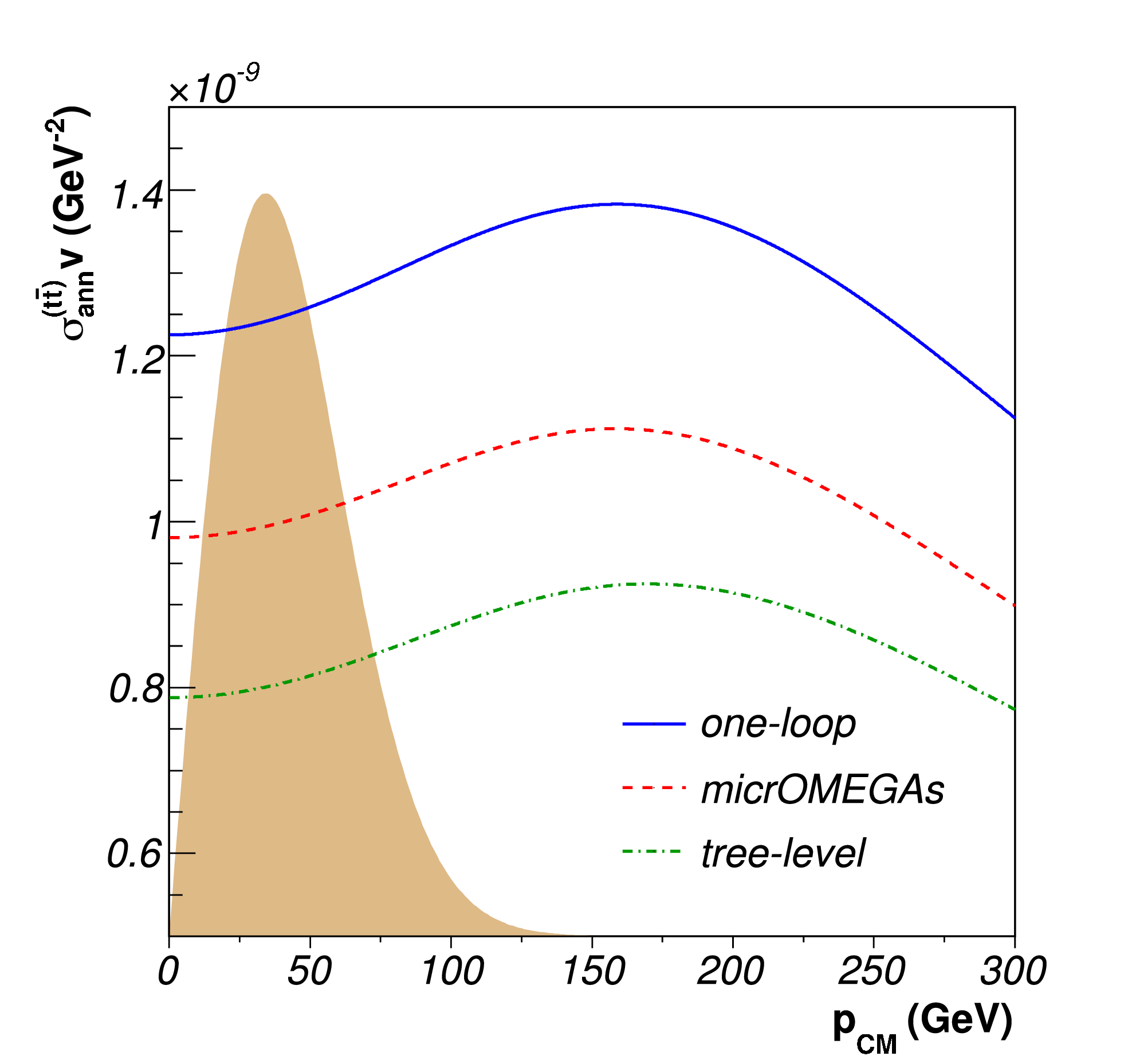}
	\end{center}
	\vspace{-5mm}
	\caption{The different contributions to the annihilation cross-section at tree-level (left) and effect of the discussed corrections (right) for a typical scenario with non-universal Higgs masses (left). The shaded area indicates the shape of the velocity distribution of the dark matter particle.}
	\label{fig:xsec}
\end{figure}

Let us now turn to the case where the annihilation proceeds mainly through the exchange of a Z-boson or a squark into a top-antitop final state. For a typical scenario within the NUHM scenario, we show in Fig.\ \ref{fig:xsec} the contributions of the different diagrams at the tree-level as well as the impact of the NLO corrections discussed in Sec.\ \ref{sec2} on the cross-section. The shaded regions indicate the velocity distribution needed for the thermal average. As can be seen, the Z- and squark-exchanges are linked by a strong destructive interference, so that the corrections to both subprocesses play an important role. At the same time, the contribution from Higgs exchange is negligible at the relevant energies. The total QCD and SUSY-QCD corrections increase the annihilation cross-section at tree-level by about 50\% \cite{DM_NUHM}. Note that the cross-section included in the public code {\tt micrOMEGAs~2.1} \cite{micrOMEGAs} takes into account parts of the NLO correction by using effective quark masses. 

\begin{figure}
	\begin{center}
	\includegraphics[scale=0.35]{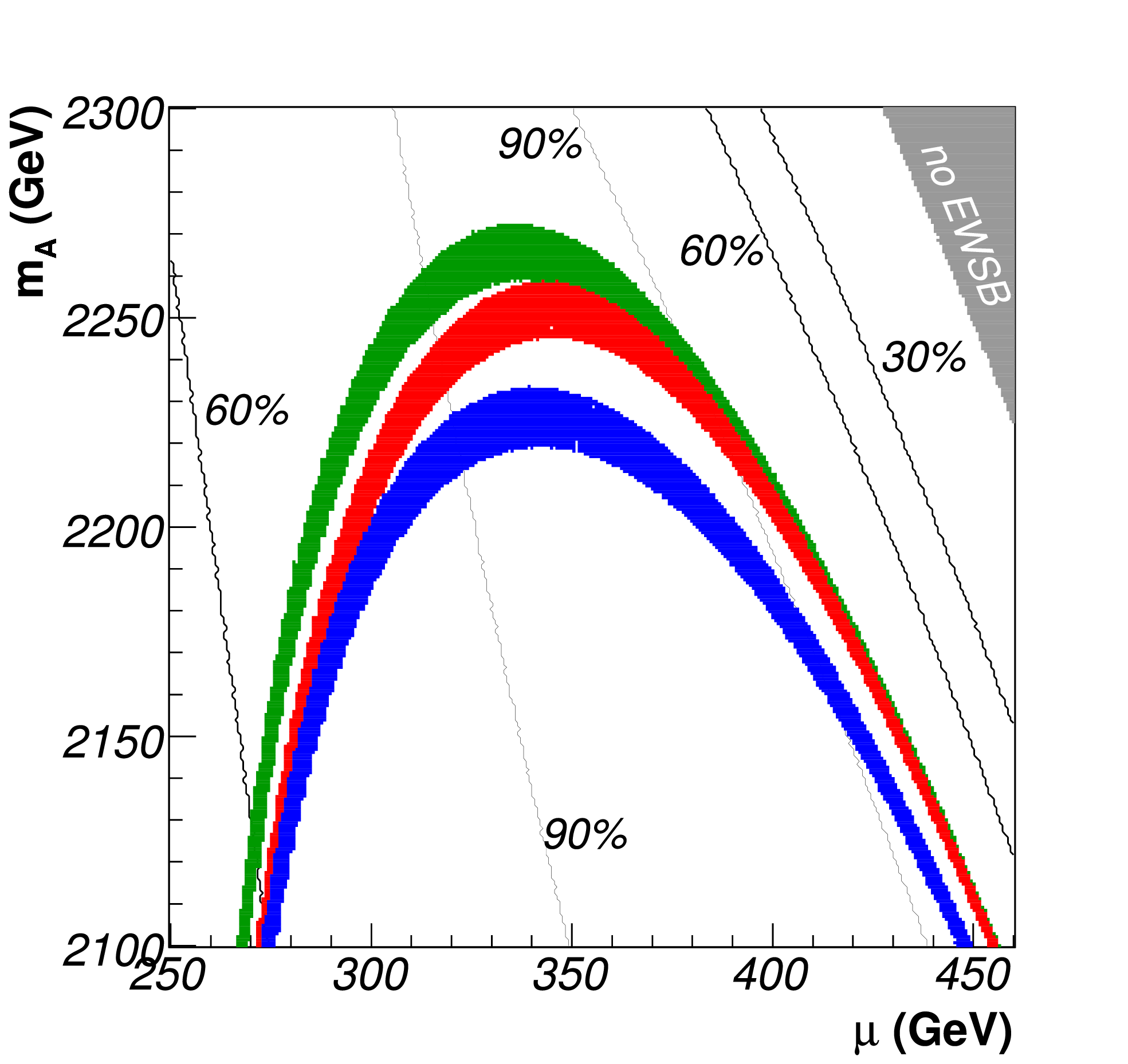}\qquad
	\includegraphics[scale=0.35]{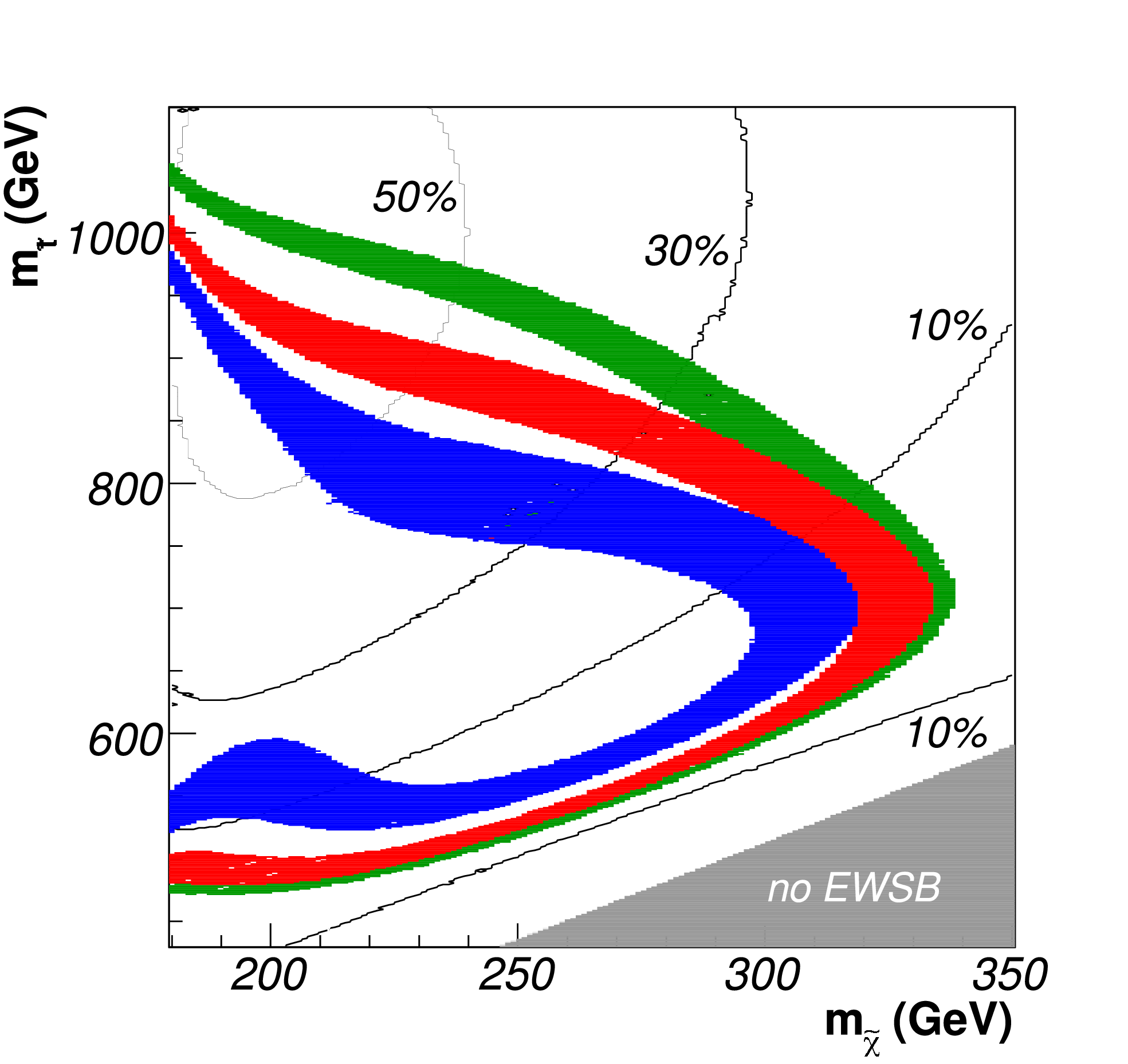}
	\end{center}
	\vspace{-5mm}
	\caption{Preferred regions in the $m_A$--$\mu$ and $m_{\tilde{t}_1}$--$m_{\tilde{\chi}_1^0}$ planes for parameter variations around two typical scenarios within non-universal Higgs and non-universal gaugino masses, respectively.}
	\label{fig:scans}
\end{figure}

Finally, in Fig.\ \ref{fig:scans} we show two examples of the preferred regions of parameter space, projected in the planes of the physical mass parameters $m_A$--$\mu$ and $m_{\tilde{t}_1}$--$m_{\tilde{\chi}^0_1}$, respectively. 
The corresponding high-scale scenarios feature non-universal Higgs masses and non-universal gaugino masses, respectively. The graphs have been obtained by scanning in $m_{H_u}$--$m_{H_d}$ and $M_1$--$m_0$ planes, respectively, and projecting the contours into the planes of physical mass parameters.
Here, the neutralino relic density has been evaluated using {\tt micrOMEGAs~2.1} \cite{micrOMEGAs}, which has been linked to the numerical package discussed in Sec.\ \ref{sec2} containing the full ${\cal O}(\alpha_s)$ corrections to the annihilation cross-section. The three bands correspond to the levels of correction discussed above and shown in Fig.\ \ref{fig:xsec}. 

As also for the A-Funnel region, the impact of the corrections is numerically more important than the uncertainty of the WMAP data. In our examples, we observe shifts of up to 50 GeV in the mass $m_A$ of the pseudoscalar Higgs boson, up to about 20 GeV for the mass $m_{\tilde{\chi}^0_1}$ of the lightest neutralino, and up to about 200 GeV for the mass $m_{\tilde{t}_1}$ of the lighter stop. Since the PLANCK satellite will deliver even more precise measurements in a very near future, the next-to-leading order corrections become even more significant.

% =======================================================================
\section{Conclusion \label{sec4}}

In summary, we have presented a complete calculation of neutralino pair annihilation into third-generation quarks at next-to-leading order in $\alpha_s$. The result has been implemented in a numerical package, which allows to evaluate their impact on the cross-section. Via a link to public tools like {\tt DarkSUSY} and {\tt micrOMEGAs} the effect of the corrections on the prediction of the relic density has been studied numerically. We have shown that the discussed QCD corrections can have a sizeable numerical impact, in particular for scenarios that are inspired by minimal supergravity such as the constrained MSSM or models with non-universal Higgs or gaugino masses \cite{AFunnel, DM_mSUGRA, DM_NUHM}. The corrections should therefore be taken into account in the extraction of supersymmetric mass parameters from cosmological data or the analysis of the parameter space of supersymmetric theories. Let us emphasize again, that both {\tt DarkSUSY} and {\tt micrOMEGAs} take into account a part of the presented corrections in terms of effective Yukawa couplings.

The obtained results can easily be generalized to the case of the annihilation of two different neutralinos or charginos into quark pairs. Moreover, annihilation involving into light quarks can be included \cite{DM_pMSSM}. While such processes are negligible within ``mSUGRA-like'' scenarios, they can play an important role in more general setups such as the 19-parameter phenomenological MSSM (pMSSM), where the soft-breaking parameters are given at the electroweak scale. 

As already stated in the Introduction, electroweak corrections can also have a sizeable impact on the prediction of the neutralino relic density, in particular for annihilation into final states with gauge bosons and leptons \cite{BaroNLO}. Further uncertainties affecting the prediction of the relic density include uncertainties on the mass spectrum of the superpartners due to renormalization group running \cite{KramlRGE} and loop-corrections to sparticle masses \cite{SPhenoNMSSM}, assumptions on the cosmological model itself \cite{Cosmo}, and uncertainties on the effective degrees of freedom due to the QCD equation of state \cite{QCD}. 

Finally, let us note that the effect of the next-to-leading order corrections discussed in this contribution are independent from the other sources of uncertainty mentioned above. Moreover, higher-order corrections to the annihilation cross-section will become even more important when the PLANCK satellite will provide more precise cosmological data in a very near future.

I would like to thank M.~Klasen and K.~Kova\v{r}\'ik for the fruitfull collaboration. 
This work is (in part) supported by the Landes-Exzellenzinitiative Hamburg.

% =======================================================================

\end{document}